\documentclass{IEEEcsmag}
\usepackage[numbers]{natbib}

\usepackage[colorlinks,urlcolor=blue,linkcolor=blue,citecolor=blue]{hyperref}
\expandafter\def\expandafter\UrlBreaks\expandafter{\UrlBreaks\do\/\do\*\do\-\do\~\do\'\do\"\do\-}
\usepackage{upmath}
\let\affil\relax
\usepackage{authblk}
\usepackage{listings}
\definecolor{IEEEblue}{RGB}{0, 112, 192}
\usepackage{booktabs}

\usepackage{algorithm}
\usepackage{algorithmicx}
\usepackage{algpseudocode}

\usepackage{amsmath}
\usepackage{amssymb}

\usepackage{microtype}

\usepackage{float}
\floatname{algorithm}{Prompt Strategy}  

\DeclareRobustCommand*{\IEEEauthorrefmark}[1]{%
    \raisebox{0pt}[0pt][0pt]{\textsuperscript{\footnotesize\ensuremath{#1}}}}
    
\begin{document}

\sptitle{Theme Article: Breakthroughs, Challenges, and Opportunities in Biological Data Visualization}

\title{AuraGenome: An LLM-Powered Framework for On-the-Fly Reusable and Scalable Circular Genome Visualizations}

\author{
    \IEEEauthorblockN{
        Chi Zhang\IEEEauthorrefmark{1}\IEEEauthorrefmark{2}\textsuperscript{a}, 
        Yu Dong\IEEEauthorrefmark{1}\textsuperscript{a}, 
        Yang Wang\IEEEauthorrefmark{1}\IEEEauthorrefmark{2}\textsuperscript{*},
        Yuetong Han\IEEEauthorrefmark{1}\IEEEauthorrefmark{2}, 
        Guihua Shan\IEEEauthorrefmark{1}\IEEEauthorrefmark{2}\IEEEauthorrefmark{3},
        and Bixia Tang\IEEEauthorrefmark{4}\IEEEauthorrefmark{5}
    }
    \IEEEauthorblockA{\IEEEauthorrefmark{1}Computer Network Information Center, Chinese Academy of Sciences, Beijing, 100083, China}
    \IEEEauthorblockA{\IEEEauthorrefmark{2}University of Chinese Academy of Sciences, Beijing, 101408, China}
    \IEEEauthorblockA{\IEEEauthorrefmark{3}Hangzhou Institute for Advanced Study, UCAS, Hangzhou, 310024, China}
    \IEEEauthorblockA{\IEEEauthorrefmark{4}National Genomics Data Center, China National Center for Bioinformation, Beijing, 100049, China}
    \IEEEauthorblockA{\IEEEauthorrefmark{5}Beijing Institute of Genomics, Chinese Academy of Sciences, Beijing, 100049, China}
    \\
    \footnotesize{\textsuperscript{a}These authors contributed equally to this work.}
    \\
     \footnotesize{\textsuperscript{*}Yang Wang is the corresponding author.}
}

\markboth{THEME ARTICLE}{THEME ARTICLE}

\begin{abstract}Circular genome visualizations are essential for exploring structural variants and gene regulation. However, existing tools often require complex scripting and manual configuration, making the process time-consuming, error-prone, and difficult to learn. To address these challenges, we introduce AuraGenome, an LLM-powered framework for rapid, reusable, and scalable generation of multi-layered circular genome visualizations. AuraGenome combines a semantic-driven multi-agent workflow with an interactive visual analytics system. The workflow employs seven specialized LLM-driven agents, each assigned distinct roles such as intent recognition, layout planning, and code generation, to transform raw genomic data into tailored visualizations. The system supports multiple coordinated views tailored for genomic data, offering ring, radial, and chord-based layouts to represent multi-layered circular genome visualizations. In addition to enabling interactions and configuration reuse, the system supports real-time refinement and high-quality report export. We validate its effectiveness through two case studies and a comprehensive user study. AuraGenome is available at: \href{https://github.com/Darius18/AuraGenome}{https://github.com/Darius18/AuraGenome}.

\end{abstract}

\maketitle

\chapteri{T}he proliferation of high-throughput sequencing technologies has transformed genomics into a data-intensive field. Modern experiments generate massive volumes of sequence-based data—such as structural variations, gene expression profiles, and chromosomal rearrangements—that are both large-scale and heterogeneous. These characteristics make data interpretation challenging using traditional visualization methods. Visualization thus plays a crucial role, helping researchers explore complex relationships, uncover patterns, and effectively communicate findings.

Circular genome visualizations offer aesthetic appeal and spatial efficiency for representing long genomic sequences and their interconnections \cite{staahlbom2024should}. Tools such as Circos \cite{circos2009} have become standard for visualizing genomic alterations and chromosome structures in circular layout. However, they rely heavily on manual configuration and scripting, imposing steep learning curves and time-consuming workflows. Researchers often need to iteratively adjust parameters to fit data-specific requirements, which can impede productivity and limit the clarity or expressiveness of the final output. These limitations highlight the demand for intelligent, automated visualization solutions tailored to the unique characteristics of circular genomic data.

Recent advances in large language models (LLMs) offer new possibilities for generating genome visualizations via natural language. While LLMs can automate complex workflows, applying them in genomics remains challenging due to the need for precise intent parsing, domain data integration, and generation of meaningful, publication-ready visualizations.

To address these challenges, we present AuraGenome, an LLM-powered framework that enables genomics experts to rapidly generate, flexibly customize, and efficiently reuse multi-layered circular genome visualizations. AuraGenome is grounded in task-specific requirements, identified through close collaboration with genomics experts to reflect real-world workflow needs. Based on the requirements, we conducted feasibility studies using LLMs for D3-based genome visualizations, and selected GPT-4o and DeepSeek-R1 as the foundation of a semantics-driven multi-agent workflow. This workflow decomposes complex visualization tasks into modular steps—such as intent recognition, layout recommendation, code generation, validation, explanation, and refinement. 

Built atop this workflow, the interactive visual analytics system supports generation and editing of circular genome visualizations in three distinct layouts—ring, radial, and chord—with multi-layer composition and domain-specific operations. A layer-aware reuse mechanism further enables tracing, adapting, and repurposing visualization steps to support efficient iteration and narrative report construction. We validate AuraGenome through two real-world case studies and both quantitative and qualitative user evaluations. In sum, the main contributions are summarized as:

\begin{enumerate}
    \item A semantics-driven multi-agent workflow that orchestrates seven LLM-based agents to transform data input into visualization-ready code output tailored for circular genome representation.
    \item A visual analytics system that enables iterative refinement, direct manipulation, and layer-aware reuse to support the construction of complex, multi-layered genome visualizations.
    \item A comprehensive evaluation through two case studies and mixed-method user studies, demonstrating the framework’s effectiveness in enhancing usability and efficiency.
\end{enumerate}

\section{Related Work}
\label{sec:related_work}

\subsection{Circular genomic visualization Applications}
\label{subsec:circular_diagrams}

Circular layouts, with their compactness and multidimensional representation capabilities, are more suitable than linear visualizations for displaying the global features of genomes. Circular layouts map data onto channels such as angles, radius, and layers, effectively illustrating hierarchical relationships and the distribution of significance across multiple data dimensions, such as gene functional regions, mutation distributions, and expression levels.

In bioinformatics, circular layouts have proven to be a highly effective visualization approach. In a study on the collinearity of the bZIP gene family in poplar, Zhao et al. \cite{zhao2021genome} employed a circular layout analysis method to explore gene functions and regulatory mechanisms by visualizing gene density, homologous blocks, and the distribution of segmental duplicated gene pairs. Peng et al. \cite{peng2022} utilized circular layouts to depict plasmid structures carrying various genes, alongside sequence alignment results of other Enterobacteriaceae bacteria sharing similar plasmid backbones, thereby illustrating plasmid gene structures, multidrug resistance regions, and homology. Garcia et al. \cite{garcia2021comprehensive}, in their study of the Lima bean genome, used circular layouts to display multiple features to understand genome organization and evolution. Similarly, Zhang et al. \cite{zhang2020genome} adopted circular layouts to present genome characteristics in tea tree research, visualizing genetic diversity and evolutionary history. Despite their widespread use, these circular layouts largely rely on manual design, limiting their efficiency and general applicability.

\subsection{Conventional Methods and Challenges in Genome Visualizations}
\label{subsec:traditional_methods}

Circos is one of the most widely used tools, generating circular visualizations through modular configurations to display genomic variations, expression levels, and chromosome structures. BioCircos \cite{cui2016biocircos} reimplements Circos using a JavaScript library, enhancing interactivity. IntelliCircos \cite{gu2025intellicircos} leverages large language models and a curated example-based dataset to support the intelligent generation of circos-based visualizations through natural language input. 

Several tools have been presented to support genome visualizations in different contexts. For example, CGView \cite{stothard2005circular} and GView \cite{petkau2010interactive} focus on browsing microbial genome annotations. BLAST Ring Image Generator \cite{alikhan2011blast} is designed specifically for comparing multiple prokaryotic genomes, utilizing BLAST results to produce circular images with an emphasis on comparative genomics. The PanVa visual analysis tool, developed by Astrid et al. \cite{van2023panva}, introduces novel aggregated graphical encoding techniques to assist researchers in exploring gene variations within complex genotype-phenotype relationships. Gosling \cite{lyi2021gosling} provides a syntactic framework supporting circular layouts, with its follow-up work, AutoGosling \cite{wang2023enabling}, leveraging deep learning to optimize genomic visualization parameters. GlyphCreator \cite{ying2021glyphcreator} is developed to parse existing circular visualization and enable secondary creation. 

These tools depend on parameterized designs, making them poorly adaptable to diverse data types and lacking support for automation. This results in low generation efficiency and a reliance on user expertise, highlighting the need for automation and intelligence, which serves as the entry point for this study.

\subsection{Visualization Generation by LLMs}
\label{subsec:large_models}

The emergence of LLMs has enabled promising advances in the automatic generation of data visualizations, greatly reducing the technical threshold for manual design. For instance, CHAT2VIS~\cite{maddigan2023chat2vis} translates natural language queries and tabular data into executable Python code; ChartGPT~\cite{tian2024chartgpt} decomposes chart generation into multi-step reasoning through fine-tuned FLAN-T5-XL; and LIDA~\cite{dibia2023lida} employs a modular pipeline to support exploratory visual analysis and narrative construction. These approaches effectively address basic visualization needs and mitigate limitations in semantic understanding and long-context reasoning.

However, the generation of complex, customized, and interactive visualizations remains a significant challenge, particularly in scientific domains such as genomics~\cite{gu2025intellicircos}. Existing models often struggle to capture the intricate data relationships, layout dependencies, and inter-track interactions that characterize these specialized charts. As noted by Cui et al.~\cite{cui2024promises}, LLMs frequently fail to retain contextual detail or integrate intermediate-level signals when processing long input sequences. In genomic visualization scenarios, these limitations lead to the omission of key features, incomplete logic, and the generation of erroneous code, especially when dealing with multi-track genome layouts or multi-omics data integration.

\begin{figure*}[htbp]
\centering
\includegraphics[width=\textwidth]{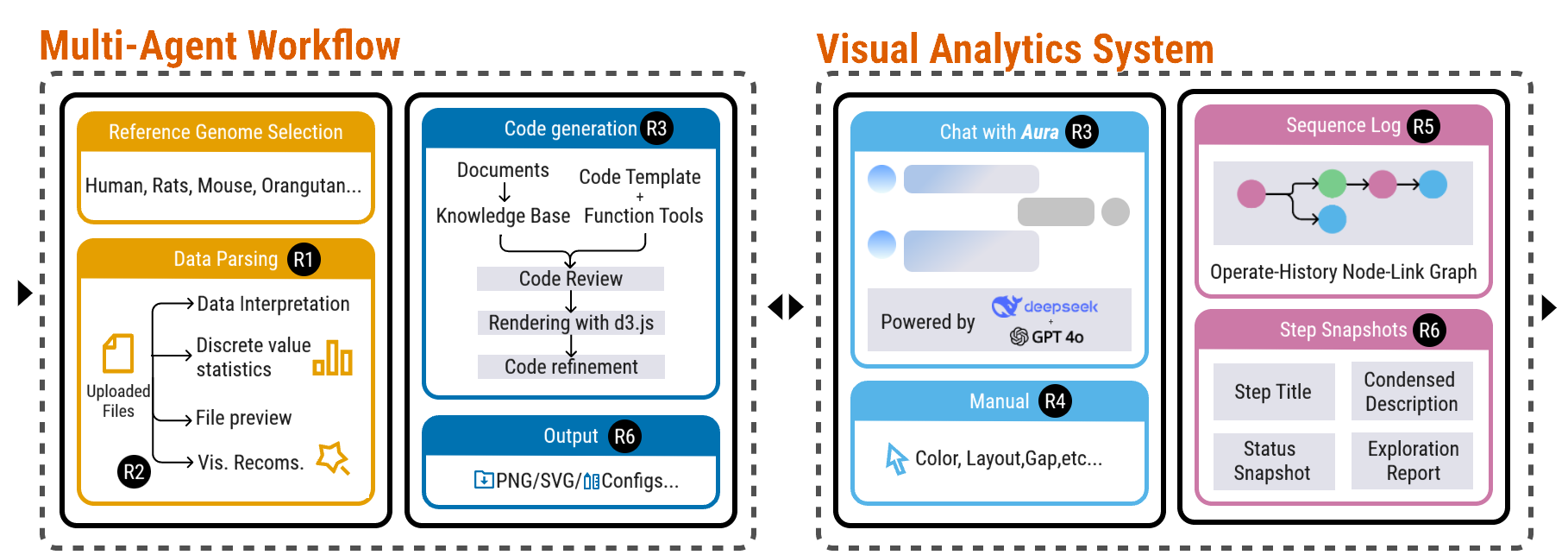}
\caption{Overview of the AuraGenome framework, which integrates a modular multi-agent workflow with an interactive visual analytics system to enable end-to-end generation, refinement, and exploration of genome-scale circular visualizations. Together, these two components fulfill six core requirements (R1–R6) to ensure scalability, interpretability, and reproducibility in genomic data visualizations.}
\label{fig:framework_architecture}
\end{figure*}

\section{Requirement Analysis}
\label{sec:requirement}

Over the past two years, we worked closely with genomics experts in ongoing research projects to identify challenges in current workflows. These collaborations revealed pressing pain points in genome visualizations and highlighted the need for more intelligent and user-friendly tools, which motivated the formulation of our research objectives.

We began by analyzing 42 genomics papers published in top-tier journals (e.g., \textit{Nature}, \textit{Cell}), focusing specifically on the use of circular visualizations to represent complex genomic information such as chromosomal rearrangements, gene interactions, and differential expression patterns. Rather than analyzing full texts, we systematically reviewed key figures in papers to identify common visualization types, usage contexts, and design goals. Through this process, we derived a taxonomy of three major layout styles: ring-based, radial, and chord-based circular genome visualizations. We also examined the tools used to generate these figures, finding that most relied on static platforms such as Circos, which offer limited customization and interactivity. These findings informed our visual design and motivated our choice of a D3-based code to support extensibility and dynamic control.

To complement these figure-derived insights with domain-specific needs, we conducted semi-structured interviews and design discussions with eight genomics experts (average experience: 8.6 years) from areas such as cancer genomics, comparative genomics, transcriptomics, and bioinformatics tool development. These experts were selected from ongoing collaborations and research networks. During sessions, experts reviewed representative circular visualizations extracted from the 42 papers, provided feedback through scenario walkthroughs, and engaged in feature prioritization and mockup evaluations. Their feedback validated the proposed layout taxonomy and further shaped our understanding of critical challenges. Based on these findings, we distilled six key requirements that a genome visual analytics framework should fulfill to enhance analytical efficiency, expressiveness, and usability.

\textbf{R1: Structured Data Ingestion and Profiling.} Efficiently ingest tabular genomic data and perform automatic feature recognition and statistical profiling to support data familiarization and downstream tasks.

\textbf{R2: Task-Aware Recommendation.} Recommend suiteable circular layouts based on semantics and analytic intent, such as detecting structural variations, comparing gene expression, or analyzing methylation patterns, thereby reducing manual trial-and-error.

\textbf{R3: Natural Language-Driven Control.} Allow users to configure visualizations using natural language commands, supporting domain-specific actions like interval binning, differential gene labeling, and feature toggling for supporting semantically aligned user interactions.

\textbf{R4: Code-Free Incremental Refinement.} Enable direct interaction with visualization styles—such as track ordering, radial spacing, and color encoding—without coding, allowing real-time, iterative refinement.

\textbf{R5: High-Flexibility Interaction.} Provide intuitive rules for code reuse, facilitating rapid adaptation to diverse visualization tasks with minimal effort.

\textbf{R6: Narrative Exploration and Summarization.} Support annotation of specific tracks or elements and generate structured summaries to facilitate reporting, interpretation, and presentation.

\section{From Requirements to AuraGenome}
\label{sec:from_requirement_to_framework}

To address the six task requirements identified through expert interviews and literature analysis, we presented AuraGenome, a framework that supports the intelligent generation and interactive refinement of circular genome visualizations. AuraGenome integrates a semantic-driven multi-agent workflow with a user-facing visual analytics system, enabling a seamless transition from raw genomic data to publication-ready visualizations. As shown in Figure~\ref{fig:framework_architecture}, AuraGenome framework consists of two core components: a semantic multi-agent workflow and a visual analytics system. 

The workflow introduces specialized LLM-based agents responsible for key subtasks, including intent recognition, data parsing (R1), visualization recommendation (R2), and code generation and refinement (R3–R5). These agents are configured based on pre-evaluation of model capabilities and are enhanced through structured prompts, modular code abstraction, and retrieval-augmented grounding.

The system enables users to refine, adjust, and manage visualizations through both direct manipulation and natural language input (R3). It supports visual parameter editing (R4), track-wise customization (R5), and summarization of analytical steps (R6). In particular, we formalize three representative layout strategies for circular genome visualizations: ring (concentric stacking of feature tracks), radial (center-out projection), and chord (arc-based linking of genomic loci). These layouts can be selected, composed, and refined interactively through novel interface elements such as the parameter panel and sequence log, which together enable visual code reuse, stepwise branching, and traceable generation history.

\section{AuraGenome Multi-Agent Workflow}
\label{sec:workflow}

\subsection{A Pre-Study for Agent Assignment}
\label{sec:prestudy_model_eval}

To construct an effective multi-agent workflow for genomic visualizations, we first conducted a pre-study evaluating the capabilities of LLMs on domain-specific tasks. We created a benchmark dataset, \textit{GenoVis-300}, containing 300 natural language prompts derived from real genomics use cases. These prompts spanned six key task types: data parsing, chart recommendation, D3.js code generation, code modification, explanation, and style adjustment. We listed three LLMs—GPT-4o-2024-11-20, GPT-o1-2024-12-27, and DeepSeek-R1-671B as condidates—to evaluate across five criteria relevant to the AuraGenome workflow: semantic understanding, code accuracy, instruction following, response efficiency, and output interpretability.

Each model was rated on a normalized 0--100 scale, where scores above 90 reflected accurate and complete results; 70--89 indicated mostly correct outputs with minor issues; 50--69 reflected partial success with notable omissions; and scores below 50 indicated critical errors. Evaluations were independently conducted by three genomics experts using anonymized outputs and a standardized rubric to ensure fairness and consistency.

\begin{figure}[h]
    \centering
    \includegraphics[width=1\linewidth]{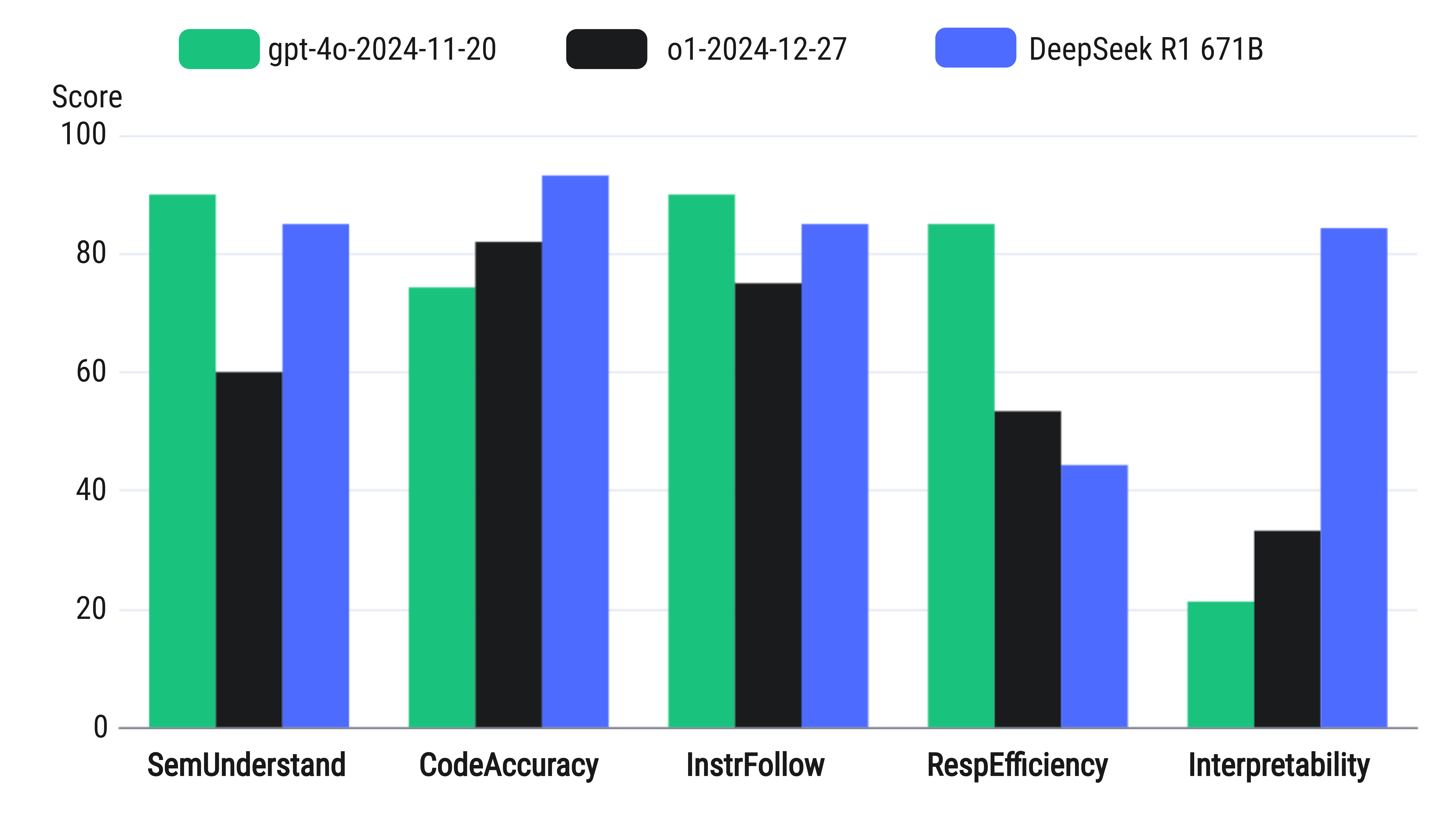}
    \caption{Comparison of LLM output assessments across five evaluation dimensions in GenoVis-300 dataset.}
    \label{fig:comparison}
\end{figure}

As shown in Figure~\ref{fig:comparison}, GPT-4o performed best in semantic parsing and explanation tasks and DeepSeek-R1 excelled in structured code generation. GPT-o1 showed moderate performance but struggled with layout consistency. These results guided our agent-model assignment strategy: GPT-4o was assigned to language-oriented agents, while DeepSeek-R1 was chosen for code-focused agents. This division supports a modular workflow where each agent operates on tasks aligned with the LLM’s strengths.

\subsection{Multi-Agent Workflow Construction}
\label{subsec:agent_workflow}

Given the diverse and complex nature of genomic visualization tasks—ranging from semantic interpretation to code synthesis and validation—monolithic prompting strategies often suffer from instability and lack of control. Based on iterative experiments on multi-agent workflow, we designed a modular, multi-agent workflow where each LLM-based agent is responsible for a specific sub-task aligned with its model capabilities. This design enables finer-grained orchestration, greater controllability, and better task transparency.

\begin{figure}[h]
    \centering
    \includegraphics[width=1\linewidth]{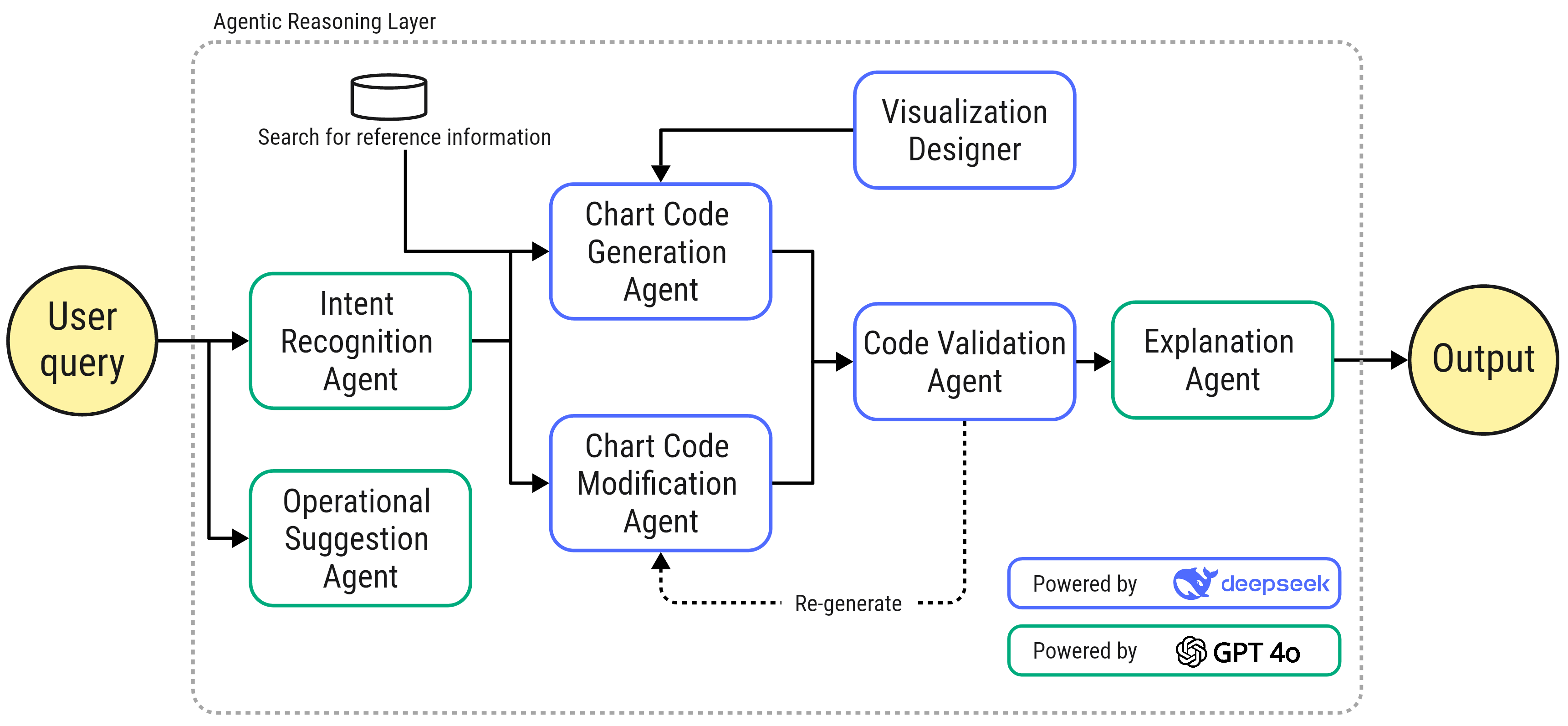}
    \caption{Overview of the multi-agent workflow for generating visualization-ready chart code from user queries. Each agent is responsible for a distinct function—ranging from intent recognition and operational suggestion to code generation, validation, and semantic explanation—while communicating through structured message passing.}
    \label{fig:multi_agent_workflow}
\end{figure}

Figure~\ref{fig:multi_agent_workflow} provides an overview of the multi-agent workflow. Each agent is positioned along a data-to-visualization pipeline, handling a specific functional scope and interfacing with others through clearly defined message-passing mechanisms. The operational details of each agent are as follows:

\textbf{1. Intent Recognition Agent.} 
The Intent Recognition Agent serves as the entry point of the multi-agent workflow, responsible for interpreting the user’s natural language input and routing it to the appropriate downstream process. Specifically, it distinguishes between two core categories of requests: visualization generation (e.g., “draw a circular layout plot for SNP distribution”) and code modification (e.g., “change the color of the mutation track”).

Rather than relying on keyword triggers or handcrafted rules, this agent leverages the LLM’s semantic parsing capability to identify user intent at a functional level. This classification enables the system to activate only the relevant agents, reducing unnecessary computation and maintaining contextual focus. By modularizing intent routing, the system supports extensibility to new task types and ensures robustness under ambiguous user input.

\textbf{2. Visualization Designer Agent.} 
The Visualization Designer Agent streamlines chart selection by analyzing the structural and semantic input—such as mutation types, zygosity profiles, and chromosomal locations—and recommending an appropriate layout.

This agent relies on embedded domain heuristics and data-profile templates to map data features to visualization strategies (e.g., heatmaps for expression data, links for structural variation). It also generates textual rationales that accompany each recommendation, improving transparency and helping users understand the rationale behind the system’s suggestions. This agent bridges the gap between raw genomic data and visualization design, enabling pre-layout planning.

\textbf{3. Chart Code Generation Agent.} 
The Chart Code Generation Agent transforms user-specified visualization designs into executable chart code. Given the complexity of genome-scale circular visualization and the flexibility of D3.js, direct code generation by LMs can be error-prone and hard to control. To address this, the agent employs a hybrid approach that integrates LLM generation with structured code templates and high-level utility functions.

This abstraction reduces the granularity of code-level decisions, allowing the agent to focus on content selection and parameter instantiation rather than low-level implementation details. It ensures that generated charts conform to best practices in layout structure and semantic consistency. The modularity also improves maintainability and supports downstream code reuse and refinement.

\textbf{4. Chart Code Modification Agent.} 
Instead of generating code from scratch, many operations involve incremental changes to existing visualizations—such as adjusting colors, track spacing, or label formats. The Chart Code Modification Agent is designed to handle such requests through in-place parameter tuning within previously generated codes.

This agent preserves the integrity of the chart structure while applying precise updates aligned with the user’s intent. It parses the code context, identifies relevant parameters, and rewrites only the affected components. This design promotes continuity in the visualization process, avoids unintended side effects, and supports real-time iteration and visual feedback.

\textbf{5. Code Validation Agent.} 
The Code Validation Agent serves as a safeguard that inspects generated code for syntax errors, data mismatches, or domain-inconsistent configurations before execution. Rather than relying solely on runtime error detection, this agent performs static checks based on domain-specific rules, such as track overlap avoidance, supported parameter values, and consistent color mapping. It uses a validation prompt schema to guide the LLM in diagnosing potential issues and suggesting fixes. While performance validation was addressed in the earlier model pre-study, this agent’s role in the pipeline is to provide real-time, explainable verification and reduce user debugging burden.

\textbf{6. Explanation Agent.} 
The Explanation Agent provides semantic annotations for generated code, helping users understand what each block of code does, why it was generated, and how it relates to the genomic dataset. This agent compares new and previous code versions, highlights the differences, and outputs human-readable commentary for each modification.

\textbf{7. Operational Suggestion Agent.} 
To support user exploration and reduce cognitive load, the Operational Suggestion Agent generates actionable prompts based on the current chart state and user goals. These suggestions include next-step operations (e.g., “add a gene density track,” “switch to cytoband view”).

\subsection{Optimization for LLM-Based Agents}
\label{subsec:agent_optimization}

While the multi-agent architecture in AuraGenome enables modular task delegation, the reliability and quality of LLM-generated outputs—especially for complex genomic visualization tasks—critically depend on how each agent is instructed, grounded, and controlled. To address this, we introduce three complementary strategies to enhance agent execution quality: a structured prompt design for role alignment, a retrieval-augmented generation mechanism for grounded code synthesis, and a modular abstraction layer to reduce code-level complexity.

\subsubsection{Retrieval-Augmented Generation for Code Generation.}
\label{subsubsec:rag_for_generation}

To further stabilize generation and align outputs with domain-specific conventions, we integrate a retrieval-augmented generation (RAG) mechanism into the chart generation and validation agents. Prior to code synthesis, the agent retrieves relevant templates from a curated knowledge base of validated chart examples, each annotated with metadata such as chart type, data modality, and genomic scale.

Retrieval is performed using a hybrid strategy that combines semantic vector matching (weight 0.7) and keyword-based exact search (weight 0.3). Retrieved examples are injected into the prompt as in-context references, grounding the model’s generation process. This approach improves fidelity to biological conventions, reduces hallucinated code patterns, and enhances output stability—especially in multi-turn interactions or layout-specific tasks.

\subsubsection{Modular Code Interfaces for Reusability and Stability.}
\label{subsubsec:modular_code_design}

Although D3.js is a flexible and expressive library for interactive visualizations, its fine-grained syntax and tightly coupled configuration logic pose significant challenges for LLMs, especially in producing robust and maintainable code. To alleviate these issues, we abstract commonly used visualization patterns into reusable utility functions and decompose the codebase into modular, semantically coherent components. This restructuring reduces the generation complexity and enables LLMs to focus on higher-level content specification rather than low-level implementation details.

The modular architecture separates core responsibilities into logically distinct layers. Data processing routines—such as genomic binning, mutation filtering, and data downsampling—are isolated from interaction behaviors, which handle operations like region highlighting, track reordering, and feature selection. The rendering logic is kept lightweight, invoking these modular components to assemble multi-layered circular layouts including ring, radial, and chord layouts. This modular architecture offers several benefits: it shortens the prompt length, reduces cognitive load for agents, and improves output generalizability across diverse tasks. Moreover, it supports structured prompt construction, allowing the system to reference and invoke specific functionality through scoped instructions.

\subsubsection{Prompt Construction Architecture for Agent Execution.}
\label{subsubsec:prompt_architecture}

To ensure consistency, reliability, and domain alignment, we design a structured prompt construction architecture supporting all LLM-based agents. Rather than relying on ad-hoc inputs, each agent dynamically assembles its prompt using a modular schema tailored to its role, input context, and output format.

This architecture significantly reduces hallucinations, improves reasoning traceability, and facilitates integration with RAG and utility libraries. The prompt template includes a role definition, task context, optionally retrieved templates, tool function references, layout constraints, and validation or output formatting blocks. This structured prompt ensures that each LLM invocation is aware of its role, operates within domain-safe constraints, and delivers well-structured, machine-consumable output. The complete construction process is summarized in Prompt Strategy~\ref{alg:prompt_construction}.

\vspace{-10pt} 

\begin{algorithm}[htbp]
\caption{Structured Prompt Construction in Blocks for Agent Execution.}
\label{alg:prompt_construction}
\begin{algorithmic}
\Require User input $U$ (e.g., data file, textual instruction); Agent role $R$; Model type $M$
\Ensure Final prompt $P$ to be executed by LLM agent
\State \textbf{Initialize} empty prompt $P$
\State Embed role definition into $P$: 
\State \hspace{1em} “You are an expert in \textit{R}, assisting the user with a genomic visualization task.”
\State Insert user context $U$ as task description
\State Add \textbf{\texttt{<search\_and\_reading>}} block:
\State \hspace{1em} Retrieve top-$k$ relevant templates from knowledge base $\mathcal{K}$ using hybrid search
\State \hspace{1em} Append results to $P$ as in-context references
\State Add \textbf{\texttt{<tool\_calling>}} block:
\State \hspace{1em} Instruct the model to use predefined utility functions from code library $\mathcal{L}$
\State Add \textbf{\texttt{<layout\_constraints>}} block:
\State \hspace{1em} Enforce chart layout logic and prevent track overlap
\If{agent $R$ is a validation agent}
    \State Add \textbf{\texttt{<error\_detection>}} block:
    \State \hspace{1em} Check for syntax, logic, and parameter errors
    \State Add \textbf{\texttt{<final\_review>}} block:
    \State \hspace{1em} Ensure modifications preserve prior valid outputs
\EndIf
\State Add \textbf{\texttt{<suggested\_adjustments>}} block:
\State \hspace{1em} Ask model to suggest up to 3 improvements (e.g., style, color, spacing)
\State Append \textbf{\texttt{<output\_format>}} block:
\State \hspace{1em} Specify required output format (e.g., JSON, JavaScript, Markdown) for system integration
\State \Return Final structured prompt $P$
\end{algorithmic}
\end{algorithm}
\vspace{-10pt}


\begin{figure*}[htbp]
    \centering
    \includegraphics[width=1\linewidth]{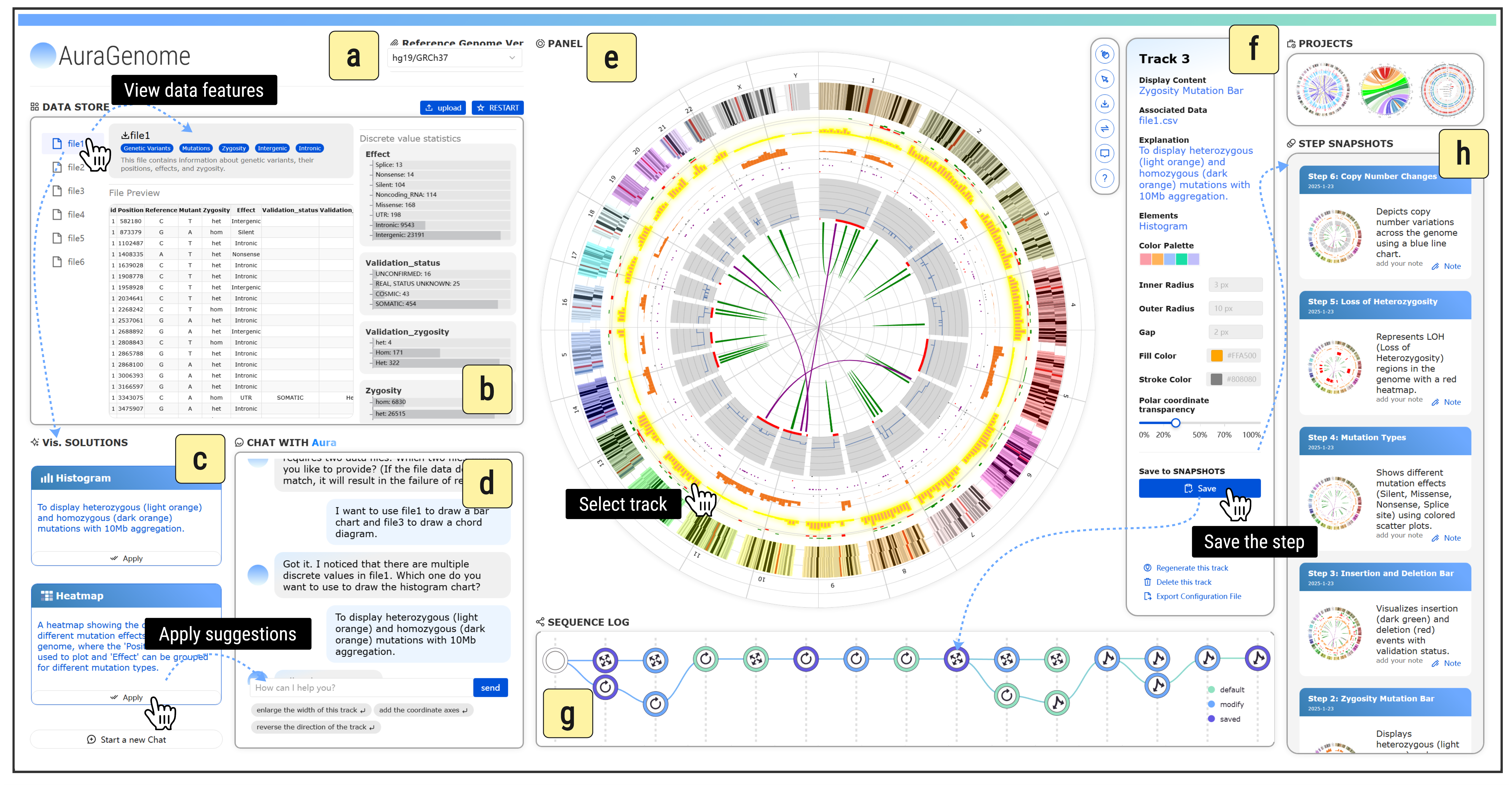}
    \caption{The visual analytics system interface of AuraGenome. Users begin by selecting reference data and exploring genomic features and statistics in (a) Reference Genome Selection and (b) Data Store View. (c) Visualization Solution Panel provides recommendations that applied and refined through (d) Chat with Aura Panel. (e) Central Visualization Panel and (f) Parameter Panel enable visual exploration and layer-wise interaction. (g) Sequence Log View captures each generation step for reuse, while (h) Project History and Step Snapshots View support project management and report output.}
    \label{fig:aura_interface}
    \vspace{-10pt}
\end{figure*}

\section{AuraGenome Visual Analytics System}
\label{sec:system_design}

\subsection{System Overview}
\label{subsec:system_overview}

The visual analytics system of AuraGenome is designed to interactively generate, refine, and manage multi-layered circular genome visualizations through a combination of natural language interaction and direct manipulation. As illustrated in Figure~\ref{fig:aura_interface}, the interface comprises multiple coordinated panels that together support the entire analytical workflow—from genome selection and data inspection to visualization generation and narrative reporting.

\textbf{(a) Reference Genome Selection}: Enables selecting from standard genome assemblies (e.g., hg19/GRCh37), which serve as the foundational coordinate system for layout construction and alignment.

\textbf{(b) Data Store View}: Presents uploaded genomic datasets alongside extracted features (e.g., mutation categories, zygosity types, and validation status). This view facilitates initial familiarization with data and informs downstream visualization recommendations.

\textbf{(c) Visualization Solution Panel}: Suggests agent-driven visualization templates based on inferred data characteristics and conventional visual mappings. Users can directly apply these suggestions to expedite layout design and reduce manual configuration.

\textbf{(d) Chat with Aura Panel}: A conversational interface that bridges users with the backend multi-agent workflow. Natural language commands—such as “increase spacing of Track 3” or “highlight LOH regions”—are semantically parsed and dispatched to the appropriate agents for contextual chart modification.

\textbf{(e) Central Visualization Panel}: Displays the multi-track circular genome chart, supporting layered rendering of structural variants, mutations, and genomic annotations with rich encoding strategies.

\textbf{(f) Track Parameter Panel}: Offers fine-grained, layer-specific customization options including radius, color, spacing, and angular transparency. All adjustments are reflected in real-time within each visualization, promoting iterative refinement.

\textbf{(g) Sequence Log View}: Visualizes the generation process through a node-link graph, where each node represents a distinct code-level output (e.g., a complete track or layout modification). In contrast to traditional action-level timelines, this abstraction emphasizes reusable configurations and logical checkpoints, allowing users to branch from prior states and efficiently explore design alternatives.

\textbf{(h) Project History and Step Snapshots View}: Maintains comprehensive project records and allows users to bookmark intermediate visual states. Snapshots can be annotated and serve as references for reporting, side-by-side comparison, or iterative editing.

The subsequent sections focus on the Central Visualization and Track Parameter Panels, along with the Sequence Log View, detailing their visual encodings and interaction design principles.

\subsection{Central Visualization and Parameter View}
\label{subsec:central}

The Central Visualization and Parameter View serves as the core canvas of the AuraGenome system, supporting dynamic construction and real-time refinement of multi-layered circular genome visualizations. As shown in Figure~\ref{fig:aura_interface}e, the visualizations are anchored to a user-specified reference genome (e.g., hg19), which defines the genomic coordinate system. New tracks are added from the outermost ring inward, following a sequential layout mechanism that enables a layered representation of heterogeneous genomic features.

The system supports three distinct track layouts—ring, radial, and chord—directly aligned with the three layout categories defined in our earlier design analysis. These styles enable users to map genomic data onto concentric rings for layered features, radial axes for attribute-specific distributions, and curved links for inter-segment relationships, respectively. A polar coordinate grid in the background provides visual scaffolding to facilitate alignment, comparison, and scale estimation. For each uploaded dataset, users can select relevant attributes (e.g., mutation zygosity, variant type, expression values) as the basis for visualizations. Through the Parameter Panel (Figure~\ref{fig:aura_interface}-e), users configure track-specific properties, including encoding types, color schemes, inner/outer radii, spacing, and border styles. Each track may include an annotation to describe the underlying data and rationale for the chosen visual encoding.

This view supports two primary interaction pathways for visualization specification: (1) recommendation-based initialization, where users adopt suggested layouts generated by the Visualization Designer Agent; and (2) dialogue-based customization, where users issue natural language commands (e.g., “make the inner track blue” or “reduce the outer radius of track 3”), which are interpreted and executed by the multi-agent workflow.

After each step, each visualization remains fully interactive. Users can refine elements directly via panel interactions, enabling real-time updates to styling, geometry, and track ordering. The interface supports reordering, in-place editing, and annotation overlays, allowing iterative refinement of the layout. Final visualizations can be exported to the Step Snapshots View, preserving the current configuration along with all parameter settings and user notes. This whole process facilitates downstream reporting and publication.

\subsection{Sequence Log View}
\label{subsec:sequence_log}

The Sequence Log View (Figure~\ref{fig:aura_interface}-f) is a node-link-based visualization designed to encode and navigate the LLM-driven process of genomic visualization generation. It explicitly represents the stepwise evolution of a user's analytical path, capturing both the structural diversity of generated charts and the decision logic behind each step.

\begin{figure}
\centering
\includegraphics[width=0.9\linewidth]{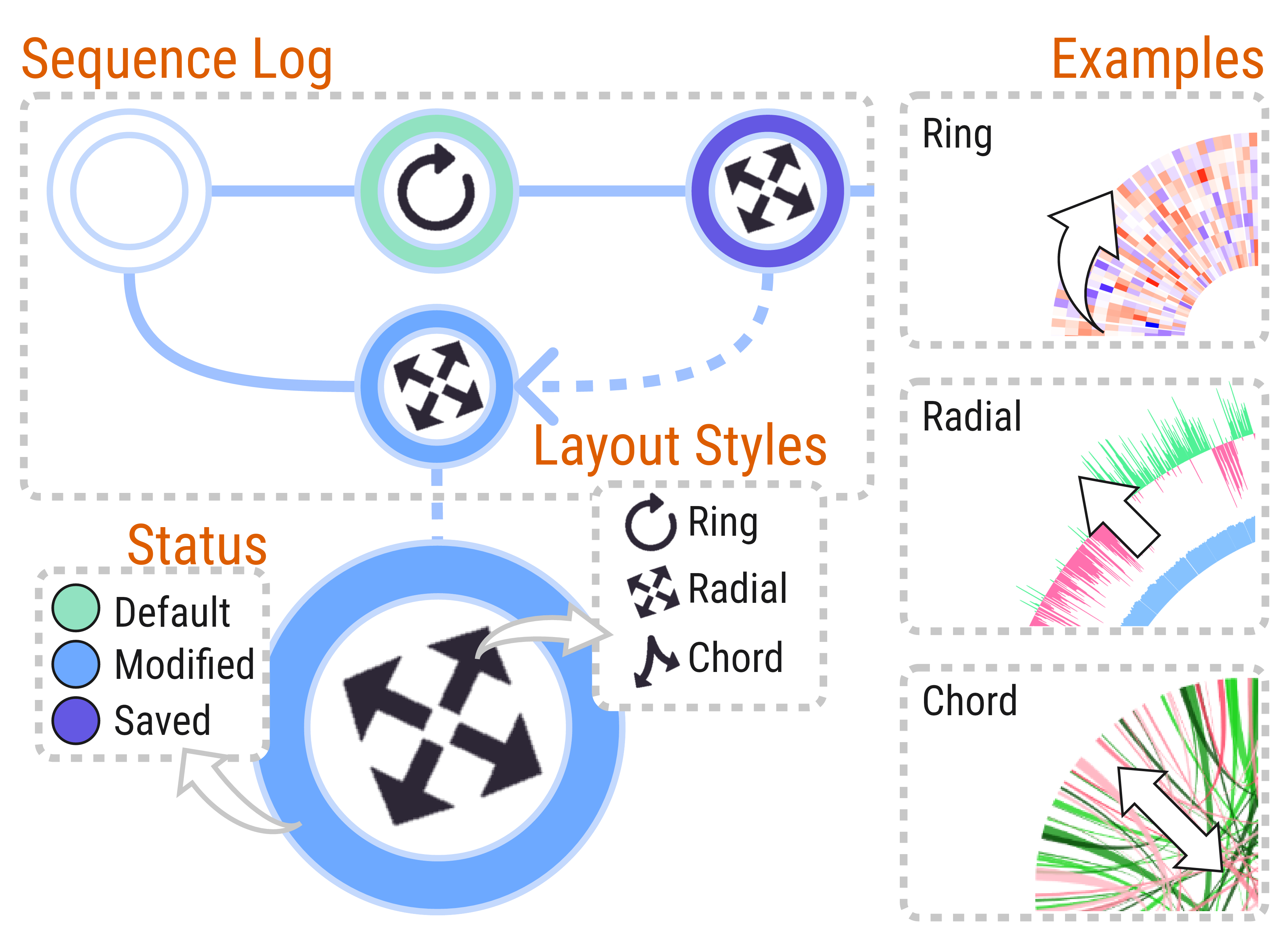}
\caption{Visual design of Sequence Log View. Each node represents a generation step, with the inner icon indicating the layout type (ring, radial, or chord) and the outer ring encoding the generation status (default, modified, or saved).}
\label{fig:visual}
\vspace{-10pt}
\end{figure}

The visual design is illustrated in Figure~\ref{fig:visual}, which details the encoding of layout types and generation statuses. Each node in the sequence is visually encoded using an inner and outer region. The inner region denotes the layout type used in that step—ring, radial, or chord—while the outer region encodes the generation status, distinguishing default, modified, and saved checkpoints through color. The sequence is arranged from left to right, with each column representing a distinct generation step. Nodes are horizontally aligned to reflect chronological order, while their vertical positions can be freely adjusted to improve visual clarity.

Connections between nodes are rendered using quadratic Bézier curves, with vertical dashed separators delineating generative stages. Each node is coupled with detailed metadata—including layout configuration, applied parameters, and associated model responses—ensuring high interpretability and workflow traceability. In cases of generation failure or suboptimal results, users can rebranch from any previous node to resume iteration with full context. The view also supports efficient reuse: users may double-click a node to draw a dashed link to subsequent steps and collaborate with agents to selectively inherit code and parameters. This helps avoid repeating past errors while maintaining focus on the current branch context.

Overall, the Sequence Log View provides a structured, interpretable, and interactive mechanism that enables users to trace generation history, modify intermediate states, and strategically reuse successful configurations, enhancing control, reliability, and efficiency in the genomic visualization process.

\section{Case Studies}
\label{sec:case_study}

To evaluate the practical effectiveness and generalizability of the AuraGenome framework, we conducted two real-world case studies involving distinct genomic analysis tasks. These studies serve to validate both the multi-agent workflow and the visual analytics system in supporting domain-specific generation, exploration, and reuse of circular genome visualizations.

The system requirements were originally derived through iterative co-design sessions with eight genomics experts. To ensure objective evaluation, the following case studies were carried out by a separate group of two extra experts—denoted as $E_A$ and $E_B$((mean experience: 10.3 years))—who were not involved in previous sections. These experts applied AuraGenome to authentic genomic analysis scenarios, enabling us to assess its usability and effectiveness in real-world contexts while minimizing evaluative bias.

\subsection{Case 1: Chromosomal Translocation and Gene Expression Profiling in AML}
\label{subsec:case1}

\begin{figure*} 
    \centering 
    \includegraphics[width=1\linewidth]{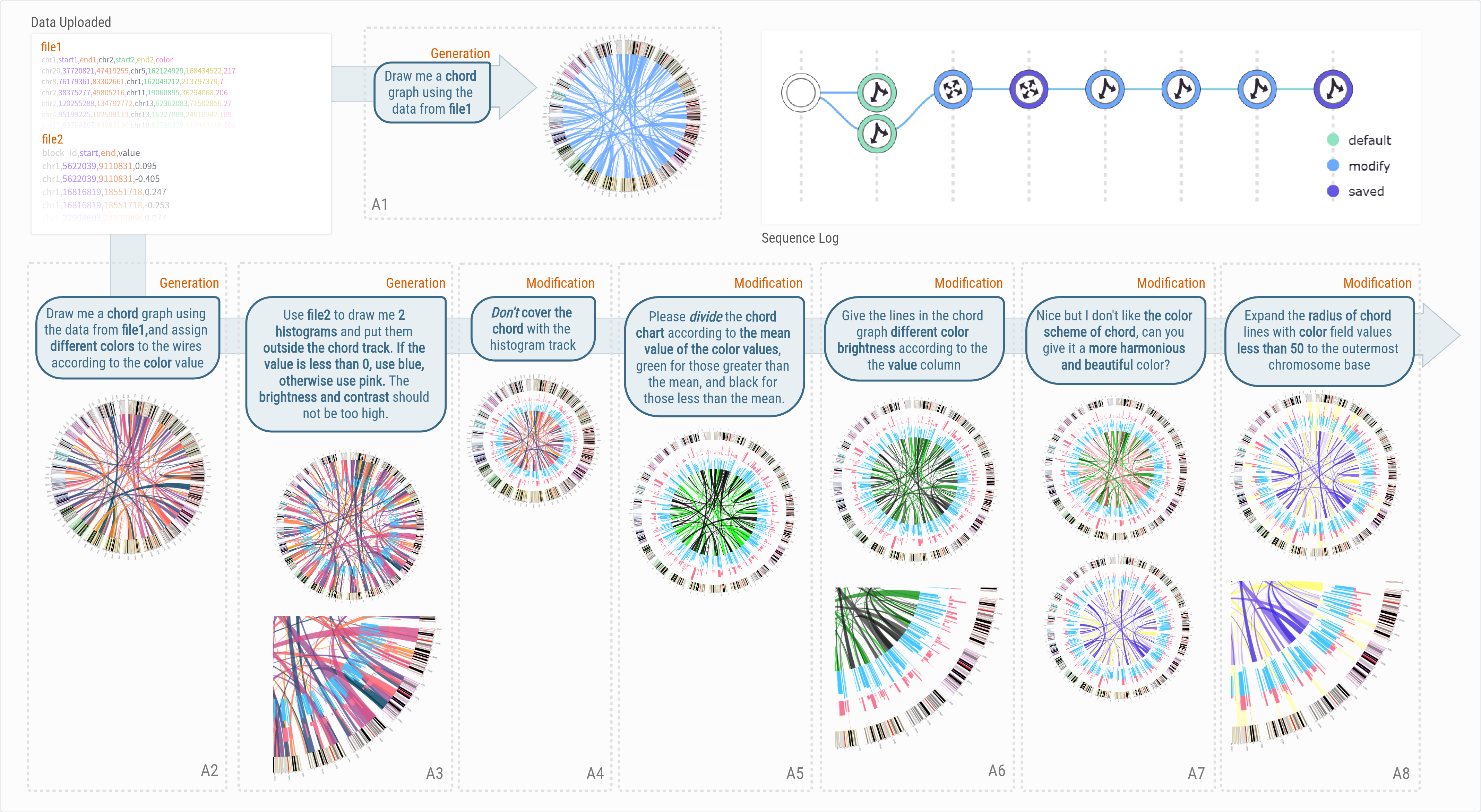} 
    \caption{Visualization workflow for AML case study. (A1-A2) Data upload and auto-parsing of chromosomal translocation and gene expression profiles. (A3) Visualization recommendation and selection of chord + ring layout. (A5) First-generation result with default color scheme. (A6-A7) Interaction-based customization of color styles for enhanced visual contrast.} 
    \label{fig:case1_workflow} 
    \vspace{-10pt}
\end{figure*}

In this case, {$E_A$} aimed to investigate chromosomal translocation patterns in acute myeloid leukemia (AML), a hematologic malignancy often driven by specific structural variants. The overall workflow is illustrated in Figure~\ref{fig:case1_workflow}.

To examine the positional distribution of breakpoints and their potential correlation with functional genomic regions, $E_A$ uploaded a curated dataset of chromosomal translocations. Leveraging AuraGenome’s automated parsing and feature extraction, the system computed translocation frequencies and mapped them to brightness levels within a circular layout (Figure~\ref{fig:case1_workflow}-A5). To visualize inter-chromosomal associations, $E_A$ selected a chord layout, which highlighted recurrent co-occurrences of breakpoints across different chromosomes.

To complement the structural data, $E_A$ incorporated a second dataset containing gene expression profiles, which was rendered as a dual-colored annular bar chart encircling the chord layout. Red bars indicated regions of upregulated expression, while blue bars represented downregulated zones. This integration enabled visual alignment of structural variants with transcriptional activity, supporting exploratory hypotheses on regulatory disruptions and disease mechanisms in AML.

When the default color scheme failed to clearly distinguish between structural and expression-related patterns, $E_A$ engaged the system’s natural language interface with the prompt: “Can you give it a more harmonious and beautiful color?” Aura responded with several palette suggestions, from which $E_A$ selected a violet and lemon-yellow combination to improve visual contrast (Figure~\ref{fig:case1_workflow}-A6, A7). The adjustment was immediately applied via the parameter panel and reinforced through backend code modifications guided by the LLM agent.

The entire process—from data upload to visualization refinement—was completed within 20 minutes. Notably, $E_A$ identified that the genomic region chr1:16816819–18551718 exhibited both a high frequency of translocation breakpoints and significant gene upregulation. This integrated insight led $E_A$ to annotate the region as a potential biomarker for AML, suggesting it as a candidate for future experimental validation.
\vspace{-10pt} 

\subsection{Case 2: Somatic Mutation Landscape of the COLO-829 Genome}
\label{subsec:case2}

To evaluate the reusability and task adaptability of AuraGenome, we conducted a second case study involving expert $E_B$, who aimed to replicate and extend insights from a widely cited study in Nature~\cite{pleasance2010comprehensive} that mapped the somatic mutation landscape of the COLO-829 melanoma genome. Building on the workflow established by $E_A$, this case demonstrates how the Sequence Log View facilitates the inheritance and contextual reuse of prior visualization logic.

\begin{figure} 
\centering 
\includegraphics[width=1\linewidth]{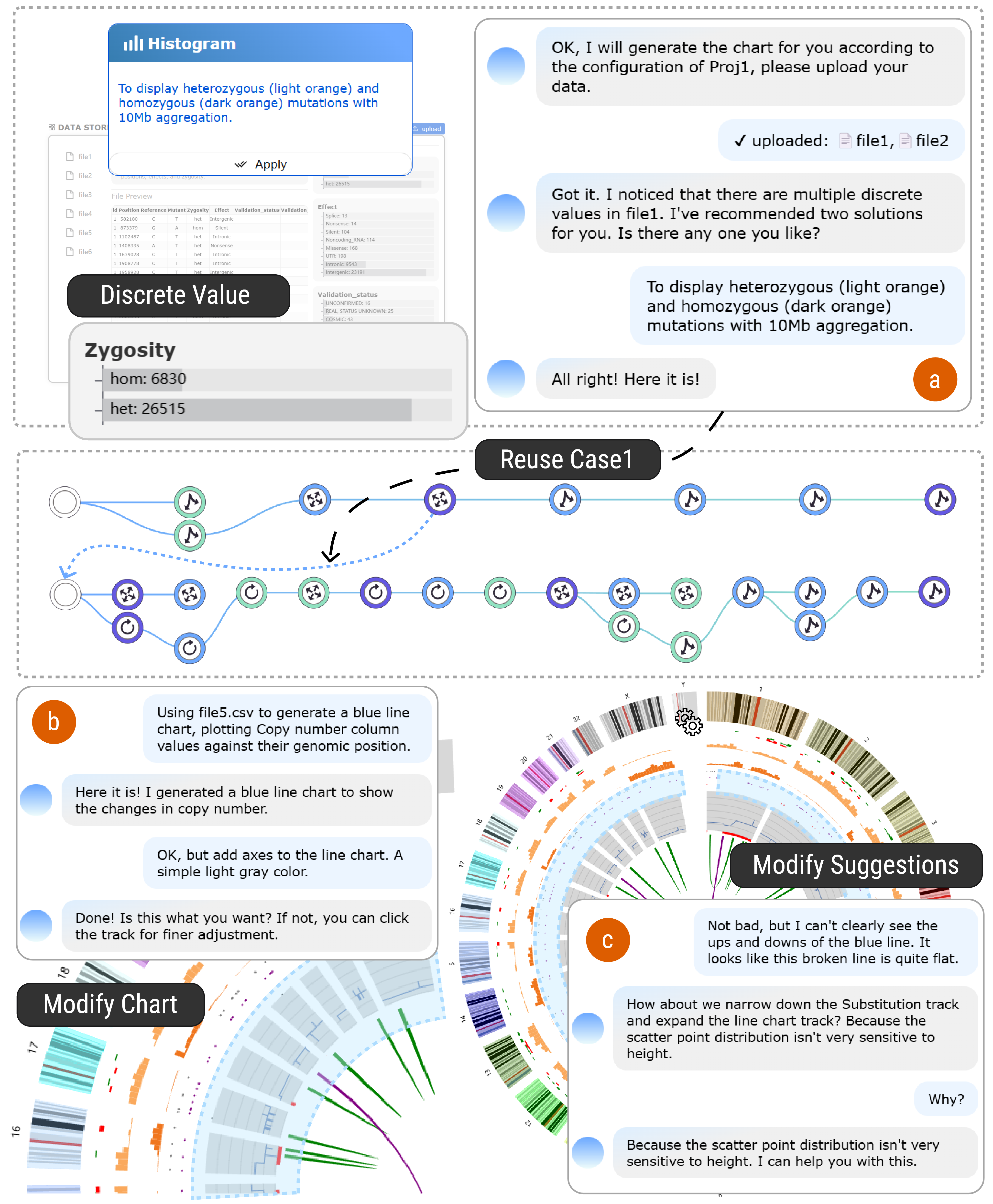} 
\caption{Visualization workflow excerpt from the COLO-829 case study. (a) Reuse the configuration from Case 1 and automatically parse the new dataset. (b) Generate and modify the copy number variation track. (c) Receive suggestions and explanations from Aura, followed by interactive and iterative modifications.} 
\label{fig:case2_workflow} 
\end{figure}

Instead of initiating a new pipeline from scratch, $E_B$ accessed the Sequence Log from Case 1 and created a new root node branching from an earlier step (Figure~\ref{fig:case2_workflow}-a). Although the visual outputs were independent, the new workflow retained structural and contextual linkage to its source, allowing $E_B$ to reuse layout strategies and code components. Upon uploading a new mutation dataset for the COLO-829 genome, AuraGenome initialized a fresh workflow using layout configurations derived from the inherited context.

Following $E_B$ uploaded the mutation dataset, AuraGenome automatically rendered two tracks: an orange bar chart encoding zygosity information and a chord layout representing chromosomal rearrangements. $E_B$ subsequently uploaded an additional dataset with insertion and deletion events and issued a natural language instruction to display validated entries using a dual-color scheme. The system accurately filtered and visualized these mutations according to the user’s intent.

Additional customization involved creating a substitution mutation track, with categorical encoding for mutation types using distinct color assignments, alongside a copy number variation (CNV) track rendered as a blue line chart. When overlapping between adjacent tracks reduced visual clarity, $E_B$ invoked the system’s recommendation module to refine the layout. Adjustments to track spacing and height were made to enhance overall readability (Figure~\ref{fig:case2_workflow}-c).

The resulting chart included seven fully customized genomic tracks and reproduced the structure of the original Nature visualizations with high fidelity. More importantly, AuraGenome’s interface preserved full interactivity throughout the process, enabling incremental edits and layout reconfigurations. The system also recorded each generation step and its associated metadata within the Step Snapshots View, producing a transparent and reproducible workflow that facilitates downstream analysis, sharing, and collaborative validation.

\section{User Study}
\label{sec:user_study}

To evaluate the effectiveness and usability of the AuraGenome, we conducted an in-lab user study focusing on system learnability, user experience across disciplines, and its utility in real analysis tasks.

\subsection{Study Design and Implementation}
\label{subsec:study_design}

\subsubsection{Participants and Tasks.}
\label{subsubsec:participants_tasks}

Twelve participants with a background in biology (average experience: 4 ± 1.5 years) were recruited for the experiment, including 10 bioinformatics researchers ($P_{1}$–$P_{10}$) and 2 wet-lab researchers ($P_{11}$–$P_{12}$). Each participant was instructed to complete the same visualization task: constructing a multi-layered circular chart using the provided genomic variation data (e.g., chromosome location, mutation type, zygosity). The task was performed using both Circos and AuraGenome. Performance was evaluated quatitative analysis through task completion time, accuracy, and a post-task questionnaire. Additional qualitative feedback was collected via follow-up interviews.

\subsubsection{Procedure.}
\label{subsubsec:procedure}

The study adopted a within-subject crossover design. Participants were randomly assigned to two groups: Group A (n = 6) used Circos first, followed by AuraGenome after a 48-hour interval; Group B (n = 6) followed the reverse order. Each session consisted of the following steps:

\textbf{1. Training (15 minutes)}: A standardized introduction to both tools was provided through documentation and guided walkthroughs.

\textbf{2. Task Execution  (maximum 150 minutes)}: We recorded the task completion time and logged all parameter modifications for later analysis.

\textbf{3. Questionnaire (10 minutes)}: Participants completed a 6-item Likert scale survey evaluating confidence, usability, and overall satisfaction.

\textbf{4. Interview (15 minutes)}: Participants provided open-ended feedback regarding system usability, task experience, and suggestions for improvement.

\subsection{Quantitative Results}
\label{subsec:quantitative_results}

\begin{figure}
    \centering
    \includegraphics[width=1\linewidth]{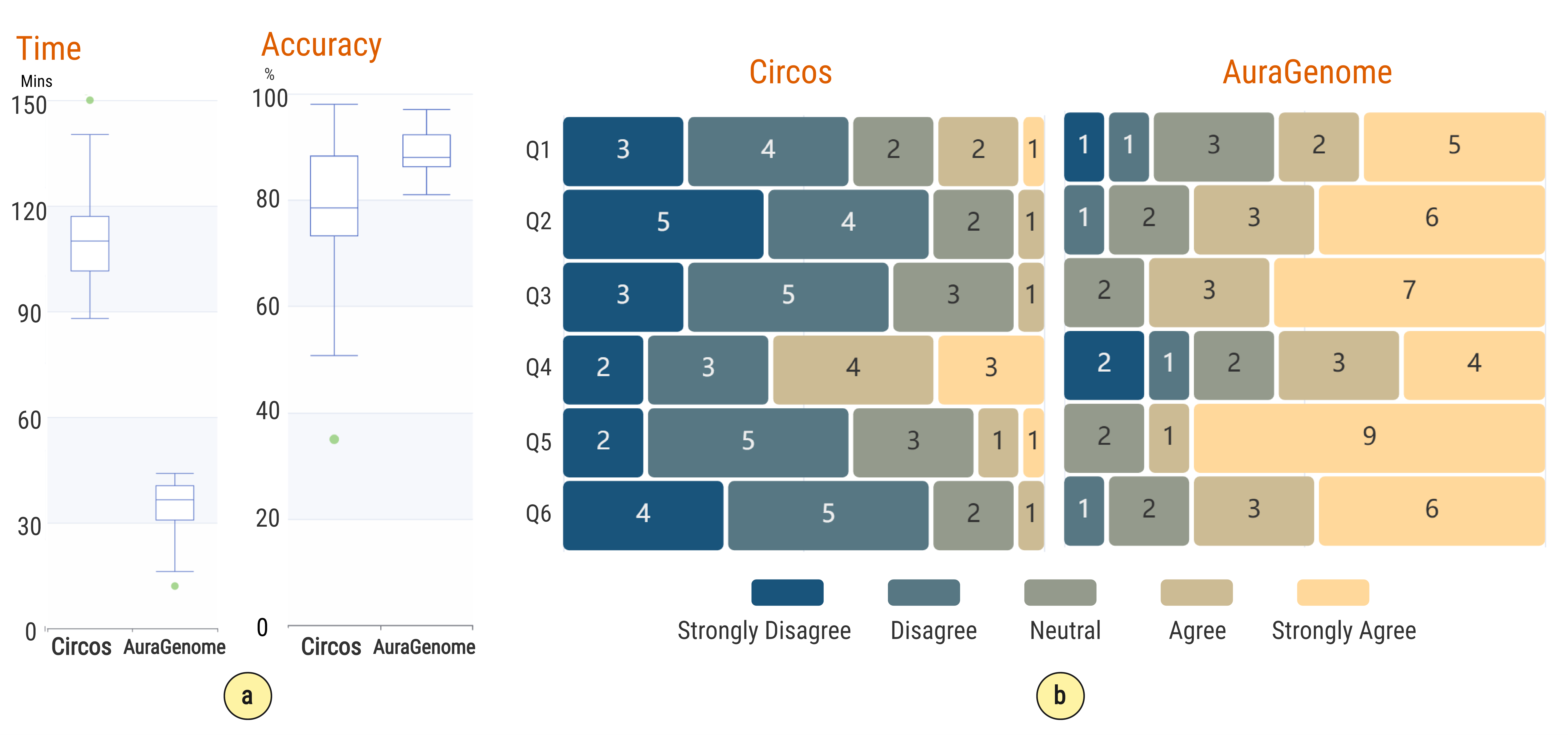}
    \caption{Quantitative analysis results: (a) Task completion time and accuracy comparison; (b) User ratings across six evaluation questions.}
    \label{fig:task_results}
\end{figure}

AuraGenome significantly outperformed Circos in both task efficiency and output accuracy. On average, participants completed the assigned visualization task in 34 minutes (SD = 9.16) using AuraGenome, compared to 112 minutes (SD = 16.5) with Circos—a 69\% reduction in time. Notably, one participant was unable to complete the task within the allotted time when using Circos. As shown in Figure~\ref{fig:task_results}-a, accuracy scores, evaluated by two independent genomics experts, averaged 89\% (SD = 4.61) with AuraGenome versus 76\% (SD = 18.31) with Circos. AuraGenome also exhibited lower performance variance across participants, suggesting improved consistency.

To assess user experience, we administered a 6-item Likert-scale questionnaire adapted from the NASA-TLX framework, covering the following dimensions (Figure~\ref{fig:task_results}-b):

\textbf{Q1}: Confidence in the accuracy of generated results.

\textbf{Q2}: Perceived ease of learning and use.

\textbf{Q3}: Efficiency in achieving intended outcomes.

\textbf{Q4}: Satisfaction with overall system functionality.

\textbf{Q5}: Ease of fine-tuning and customization.

\textbf{Q6}: Overall satisfaction and enjoyment.

AuraGenome achieved significantly higher ratings than Circos on five of six questions (Q1–Q3, Q5–Q6; $p < 0.05$, Mann–Whitney U test), with Q4 (functionality) also trending positively ($p = 0.0588$). Additionally, 11 out of 12 participants reported that AuraGenome was easier to use and customize, and 8 expressed greater confidence in the correctness.

These findings demonstrate that AuraGenome provides notable improvements in usability, efficiency, and accuracy over traditional tools, particularly benefiting users with limited programming expertise.

\subsection{Qualitative User Reflections}
\label{subsec:qualitative_feedback}

To complement the quantitative results, we collected open-ended feedback from 12 participants through post-task interviews and questionnaires. We summarized them for futher thematic analysis. The result revealed four recurring reflections about AuraGenome’s usability, functionality, and areas for growth.

\subsubsection{Natural Language as a Shift in Interaction Paradigm.}

All participants embraced the natural language interface as a major improvement over traditional config-file editing. Chat with Aura lowered the barrier to entry and allowed users to express visualization intent more directly ($P_{2}$, $P_{3}$, $P_{8}$, $P_{10}$). The real-time visualization generation further reduced trial-and-error, streamlining the configuration process ($P_{1}$, $P_{2}$, $P_{4}$, $P_{9}$).

\subsubsection{Integration of Domain Knowledge.}

Participants highlighted that AuraGenome reduced the burden of preselecting visualization strategies by automatically recommending composite charts tailored to the input data type. In contrast to traditional workflows requiring prior layout selection before track configuration. $P_{6}$ noted, \textit{“The ‘variant + heatmap’ combo matched my study better than my own plan.”} By surfacing context-aware suggestions, the system not only accelerated chart creation but also introduced users to more informative visual encodings than they might have initially decided.

\subsubsection{Cross-Disciplinary Usability.}

The system was seen as accessible to both computational and wet-lab users. While experienced users valued the ability to issue fine-grained commands, including suggestions for direct SVG editing ($P_{4}$, $P_{10}$), wet-lab participants noted that they could independently produce publication-ready charts—an empowering capability for non-programmers ($P_{11}$, $P_{12}$). This cross-disciplinary flexibility was highlighted as a key advantage over traditional tools.

\subsubsection{Areas for Growth.}

Few participants also identified opportunities for refinement, particularly in balancing automation with control. Some expressed a desire for richer data previews ($P_{1}$), editable annotation history ($P_{4}$), and better visibility into the sequence of steps taken. Others ($P_{5}$) noted that automated defaults occasionally misaligned with their intent, suggesting that optional confirmation or stepwise override could improve usability.

\vspace{-10pt}

\section{Discussion and Limitations}
\label{subsec:discussion_limitations}

\subsection{Capability of AuraGenome in Supporting Complex Visualization Tasks}
AuraGenome's multi-agent workflow is deliberately designed to decompose complex visualization requests into manageable sub-tasks—such as intent recognition, layout planning, and code generation—handled by role-specific LLM agents. This structured collaboration enables the system to reliably interpret domain-specific goals and produce customized, multi-track genome visualizations. Unlike prior approaches that generate Circos-based configuration or rely on predefined templates, AuraGenome produces flexible D3-based script, supporting rich interactivity, dynamic interaction, and reuse. Together, these design elements make AuraGenome well-suited for handling the complex tasks of circular genome visualizations.

\subsection{End-User Expertise and Interaction Fluency}
User experience varied with domain expertise, yet AuraGenome proved adaptable across the spectrum—from bioinformatics specialists to experimental researchers with limited visualization backgrounds. The natural language interface, paired with real-time visual feedback and well-aligned agent behavior, enabled users to express high-level goals without needing technical vocabulary or implementation knowledge. Participants consistently highlighted the system’s predefined interactive mechanisms—such as dynamic track reordering, attribute refinement, and visual reuse—as unexpectedly intuitive and powerful. These features not only facilitated iterative correction and fine-tuning, but also encouraged exploratory behavior, allowing users to experiment with alternative layout patterns, highlight strategies, or comparative track compositions. AuraGenome’s layered feedback, affordance-rich UI, and modular agent structure collectively support both directed editing and open-ended exploration, reinforcing its value as a user-centric tool.

\subsection{Limitations and Boundary Conditions}
While AuraGenome demonstrates strong adaptability, several functional boundaries remain. The current workflow and system is tailored for circular genome visualizations and does not yet support linear or hybrid visualization generation, which are the same significant in genomic analysis. Additionally, performance would be modestly affected by the agents when handling ambiguous, conflicting, or poorly structured prompts, occasionally leading to redundant tracks or minor layout inconsistencies. However, these issues do not interrupt task execution. Thanks to the multi-agent workflow, such errors are mitigated through agent-level coordination and fallback strategies, ensuring smooth task progression. These observations point to directions for further refinement, while confirming that AuraGenome remains robust and reliable in handling complex genomic visualization tasks.

\section{Conclusion and Future Work}
\label{sec:conclusion}
In this paper, we presented AuraGenome, an LLM-powered framework for generating rapid, reusable, and scalable circular genome visualizations. By integrating multi-agent collaboration with interactive visual analytics, AuraGenome enables natural language-driven visualization generation, significantly improving both efficiency and output quality. Two real-world case studies and a comprehensive user study demonstrate its effectiveness in enhancing task performance and reducing cognitive load in genomic research workflows.

In future work, we plan to refine the multi-agent workflow to capture more fine-grained user intents and support richer interactions with visualization outcomes. We also aim to extend support beyond circular layouts, enabling multi-view genomic representations that accommodate diverse analytical needs. Additionally, we will enhance the robustness of LLM-driven generation through improved retrieval grounding and validation strategies, thereby increasing reliability and reducing the risk of hallucinated outputs.

\section{ACKNOWLEDGMENTS}
This work was supported by STI2030-Major Projects No.2021ZD0200200 and Beijing Natural Science Foundation No.4254090.

\def\refname{REFERENCES}

\bibliographystyle{IEEEtran}

\begin{thebibliography}{18}

\bibitem{circos2009}
M.~Krzywinski, J.~Schein, I.~Birol, J.~Connors, R.~Gascoyne, D.~Horsman, S.~J. Jones, and M.~A. Marra, ``Circos: an information aesthetic for comparative genomics,'' \emph{Genome Research}, vol.~19, no.~9, pp. 1639--1645, 2009.

\bibitem{cui2024promises}
Y.~Cui, W.~G. Lily, Y.~Ding, L.~Harrison, F.~Yang, and M.~Kay, ``Promises and pitfalls: using large language models to generate visualization items,'' \emph{IEEE Transactions on Visualization and Computer Graphics}, 2024.

\bibitem{zhao2021genome}
K.~Zhao, S.~Chen, W.~Yao, Z.~Cheng, B.~Zhou, and T.~Jiang, ``Genome-wide analysis and expression profile of the bZIP gene family in poplar,'' \emph{BMC Plant Biology}, vol.~21, pp. 1--16, 2021.

\bibitem{peng2022}
Z.~Peng, Z.~Hu, Z.~Li, X.~Zhang, C.~Jia, T.~Li, M.~Dai, C.~Tan, Z.~Xu, B.~Wu \emph{et~al.}, ``Antimicrobial resistance and population genomics of multidrug-resistant \emph{Escherichia coli} in pig farms in mainland China,'' \emph{Nature Communications}, vol.~13, no.~1, p. 1116, 2022.

\bibitem{garcia2021comprehensive}
T.~Garcia, J.~Duitama, S.~S. Zullo, J.~Gil, A.~Ariani, S.~Dohle, A.~Palkovic, P.~Skeen, C.~I. Bermudez-Santana, D.~G. Debouck \emph{et~al.}, ``Comprehensive genomic resources related to domestication and crop improvement traits in Lima bean,'' \emph{Nature Communications}, vol.~12, no.~1, p. 702, 2021.

\bibitem{zhang2020genome}
W.~Zhang, Y.~Zhang, H.~Qiu, Y.~Guo, H.~Wan, X.~Zhang, F.~Scossa, S.~Alseekh, Q.~Zhang, P.~Wang \emph{et~al.}, ``Genome assembly of wild tea tree DASZ reveals pedigree and selection history of tea varieties,'' \emph{Nature Communications}, vol.~11, no.~1, p. 3719, 2020.

\bibitem{cui2016biocircos}
Y.~Cui, X.~Chen, H.~Luo, Z.~Fan, J.~Luo, S.~He, H.~Yue, P.~Zhang, and R.~Chen, ``BioCircos. js: an interactive Circos JavaScript library for biological data visualization on web applications,'' \emph{Bioinformatics}, vol.~32, no.~11, pp. 1740--1742, 2016.

\bibitem{stothard2005circular}
P.~Stothard and D.~S. Wishart, ``Circular genome visualization and exploration using CGView,'' \emph{Bioinformatics}, vol.~21, no.~4, pp. 537--539, 2005.

\bibitem{petkau2010interactive}
A.~Petkau, M.~Stuart-Edwards, P.~Stothard, and G.~Van~Domselaar, ``Interactive microbial genome visualization with GView,'' \emph{Bioinformatics}, vol.~26, no.~24, pp. 3125--3126, 2010.

\bibitem{alikhan2011blast}
N.-F. Alikhan, N.~K. Petty, N.~L. Ben~Zakour, and S.~A. Beatson, ``BLAST Ring Image Generator (BRIG): simple prokaryote genome comparisons,'' \emph{BMC Genomics}, vol.~12, pp. 1--10, 2011.

\bibitem{van2023panva}
A.~van~den Brandt, E.~M. Jonkheer, D.-J.~M. van Workum, H.~van~de Wetering, S.~Smit, and A.~Vilanova, ``PanVA: pangenomic variant analysis,'' \emph{IEEE Transactions on Visualization and Computer Graphics}, 2023.

\bibitem{lyi2021gosling}
S.~LYi, Q.~Wang, F.~Lekschas, and N.~Gehlenborg, ``Gosling: a grammar-based toolkit for scalable and interactive genomics data visualization,'' \emph{IEEE Transactions on Visualization and Computer Graphics}, vol.~28, no.~1, pp. 140--150, 2021.

\bibitem{wang2023enabling}
Q.~Wang, X.~Liu, M.~Q. Liang, S.~L’Yi, and N.~Gehlenborg, ``Enabling multimodal user interactions for genomics visualization creation,'' in \emph{2023 IEEE Visualization and Visual Analytics (VIS)}, 2023, pp. 111--115.

\bibitem{ying2021glyphcreator}
L.~Ying, T.~Tang, Y.~Luo, L.~Shen, X.~Xie, L.~Yu, and Y.~Wu, ``GlyphCreator: towards example-based automatic generation of circular glyphs,'' \emph{IEEE Transactions on Visualization and Computer Graphics}, vol.~28, no.~1, pp. 400--410, 2021.

\bibitem{maddigan2023chat2vis}
P.~Maddigan and T.~Susnjak, ``Chat2vis: generating data visualizations via natural language using chatgpt, codex and gpt-3 large language models,'' \emph{IEEE Access}, vol.~11, pp. 45181--45193, 2023.

\bibitem{tian2024chartgpt}
Y.~Tian, W.~Cui, D.~Deng, X.~Yi, Y.~Yang, H.~Zhang, and Y.~Wu, ``Chartgpt: leveraging llms to generate charts from abstract natural language,'' \emph{IEEE Transactions on Visualization and Computer Graphics}, 2024.

\bibitem{dibia2023lida}
V.~Dibia, ``LIDA: a tool for automatic generation of grammar-agnostic visualizations and infographics using large language models,'' \emph{arXiv preprint arXiv:2303.02927}, 2023.

\bibitem{pleasance2010comprehensive}
E.~D. Pleasance, R.~K. Cheetham, P.~J. Stephens, D.~J. McBride, S.~J. Humphray, C.~D. Greenman, I.~Varela, M.-L. Lin, G.~R. Ord{\'o}{\~n}ez, G.~R. Bignell \emph{et~al.}, ``A comprehensive catalogue of somatic mutations from a human cancer genome,'' \emph{Nature}, vol.~463, no. 7278, pp. 191--196, 2010.

\bibitem{staahlbom2024should}
E.~St{\aa}hlbom, J.~Molin, A.~Ynnerman, and C.~Lundstr{\"o}m, ``Should I make it round? Suitability of circular and linear layouts for comparative tasks with matrix and connective data,'' in \emph{Computer Graphics Forum}, vol.~43, no.~3, 2024, p. e15102.

\bibitem{gu2025intellicircos}
M.~Gu, J.~Zhu, Q.~Wang, F.~Wang, X.~Wen, Y.~Wang, and M.~Zhu, ``IntelliCircos: a data-driven and AI-powered authoring tool for Circos plots,'' \emph{arXiv preprint arXiv:2503.24021}, 2025.

\end{thebibliography}

\begin{IEEEbiography}{{\textcolor{IEEEblue}{Chi Zhang}}} is student at Computer Network Information Center, Chinese Academy of Sciences at Beijing, China. Contact him at zhangc@cnic.cn. 
\end{IEEEbiography}

\begin{IEEEbiography}{{\textcolor{IEEEblue}{Yu Dong}}} is assistant professor at Computer Network Information Center, Chinese Academy of Sciences at Beijing, China. Contact him at dongyu@cnic.cn.
\end{IEEEbiography}

\begin{IEEEbiography}{{\textcolor{IEEEblue}{Yang Wang}}} is associate professor at Computer Network Information Center, Chinese Academy of Sciences at Beijing, China. Contact him at wangyang@sccas.cn.
\end{IEEEbiography}

\begin{IEEEbiography}{{\textcolor{IEEEblue}{Yuetong Han}}} is student at Computer Network Information Center, Chinese Academy of Sciences at Beijing, China. Contact her at ythan@cnic.cn. 
\end{IEEEbiography}

\begin{IEEEbiography}{{\textcolor{IEEEblue}{Guihua Shan}}} is professor at Computer Network Information Center, Chinese Academy of Sciences at Beijing, China. Contact her at sgh@cnic.cn.
\end{IEEEbiography}

\begin{IEEEbiography}{{\textcolor{IEEEblue}{Bixia Tang}}} is associate professor at China National Center for Bioinformation at Beijing, China. Contact her at tangbx@big.ac.cn.
\end{IEEEbiography}

\end{document}